\documentclass[twocolumn,prd, superscriptaddress, showpacs]{revtex4-1}
\usepackage{hyperref,xcolor}
\usepackage{amsmath,amssymb}
\usepackage{graphicx}
\usepackage{xcolor}
\usepackage{tikz}
\usepackage{bm}
\usepackage{slashed}
\usepackage{placeins}

\newcommand{\Z}{\ensuremath{\mathbb{Z}}}

\newcommand{\rd}{\ensuremath{\mathrm{d}}}

\providecommand{\abs}[1]{\left\lvert#1\right\rvert}
\providecommand{\norm}[1]{\left\lVert#1\right\rVert}
\newcommand{\expv}[1]{\left\langle#1\right\rangle}

\newcommand{\bphi}{\bm{\Phi}}
\newcommand{\bdphi}{\bm{\delta\Phi}}

\newcommand{\be}{\begin{equation}}
\newcommand{\ee}{\end{equation}}
\newcommand{\benn}{\nonumber\begin{equation}}
\newcommand{\eenn}{\nonumber\end{equation}}
\def\bea{\begin{eqnarray}} \def\eea{\end{eqnarray}}
\def\beann{\begin{eqnarray*}} \def\eeann{\end{eqnarray*}}
\def\lsim{\raise0.3ex\hbox{$<$\kern-0.75em\raise-1.1ex\hbox{$\sim$}}}
\def\gsim{\raise0.3ex\hbox{$>$\kern-0.75em\raise-1.1ex\hbox{$\sim$}}}

\newcommand*\xbar[1]{%
  \hbox{%
    \vbox{%
      \hrule height 0.5pt % The actual bar
      \kern0.3ex%         % Distance between bar and symbol
      \hbox{%
        \kern-0.1em%      % Shortening on the left side
        \ensuremath{#1}%
        \kern-0.1em%      % Shortening on the right side
      }%
    }%
  }%
}

\begin{document}

\title{
Higgs-Yukawa model with higher dimension operators via extended mean field theory}
\date{\today}
\author{Oscar Akerlund}
\affiliation{Institut f\"ur Theoretische Physik, ETH Zurich, CH-8093 Z\"urich, Switzerland}
\affiliation{Kavli Institute for Theoretical Physics, University of California,
Santa Barbara, CA 93106-4030 USA}
\author{Philippe de Forcrand}
\affiliation{Institut f\"ur Theoretische Physik, ETH Zurich, CH-8093 Z\"urich, Switzerland}
\affiliation{Kavli Institute for Theoretical Physics, University of California,
Santa Barbara, California 93106-4030 USA}
\affiliation{CERN, Physics Department, TH Unit, CH-1211 Geneva 23, Switzerland}

\begin{abstract}
Using extended mean field theory (EMFT) on the lattice, we study properties of the Higgs-Yukawa model as an approximation of
the standard model Higgs sector, and the effect of higher dimension operators.
We remark, as has been noted before, that the discussion of vacuum stability is completely modified in the presence
of a $\phi^6$ term, and that the Higgs mass no longer appears fine tuned.
We also study the finite temperature transition. Without higher dimension operators the
transition is found to be second order (crossover with gauge fields) for the experimental value of the Higgs mass $M_h=125$ GeV.
By taking a $\phi^6$ interaction in the Higgs potential as a proxy for a UV completion of the standard model, the transition becomes
stronger and turns first order if the scale of new physics, i.e. the mass of the lightest mediator particle, is around $1.5$ TeV.
This implies that electroweak baryogenesis may be viable in models which introduce new particles around that scale.
\end{abstract}

\maketitle

\section{Introduction}\label{sec:introduction}\noindent
\begin{tikzpicture}[overlay, remember picture]
\path (current page.north east) ++(-1,-0.5) node[below left] {\texttt{\footnotesize CERN-PH-TH/2015-210}}
++(0,-0.3) node[below left]{\texttt{\footnotesize NSF-KITP-15-123}};
\end{tikzpicture}
Even before the Higgs boson was discovered several studies where the standard model (SM) couplings were run to very large
energies were conducted. This resulted in both upper and lower bounds on the Higgs mass stemming from the triviality of the
Higgs self-interaction and the electroweak (EW) vacuum stability, respectively. Sandwiched between these bounds is a small
region of Higgs masses for which the SM can be run at least up to the Planck scale~\cite{Espinosa:1995se,Branchina:2014rva}.
Due to the huge success of the SM when
it comes to explaining results from accelerator experiments it was predicted that the Higgs mass would indeed lie inside this
region and that the UV completion of the SM would enter only at the Planck scale where gravity becomes important. As the Atlas and
CMS experiments announced the discovery of a Higgs-like particle at $125$ GeV, which is at least very close to the special region,
these speculations about no new physics before the Planck scale got a lot of deserved attention. It is certainly interesting and
important to thoroughly investigate this possibility, see~\cite{Espinosa:1995se} and references therein.

However, the apparent special value of the Higgs mass of course in no way excludes new physics at a lower scale and we will in
this paper deal with generic aspects of UV completions at a fairly low scale of a few to tens of TeV. So let us depart from the
arbitrarily postulated quartic self-interaction of the Higgs field. Since it does not include dark matter or gravity,
the SM is only an effective theory anyway, and there is no reason to assume a renormalizable Higgs sector.
We may add higher-dimension operators, which may not directly be relevant for the low energy physics but which
can, for example, play an important role for the stability of the EW vacuum or in the context of EW baryogenesis, in an effective
field theory way. This will of course drastically change the running of the SM coupling constants so the question of whether the
Higgs mass puts the universe in a near critical state or not loses its relevance in this context. We choose to consider generic
higher-dimension operators instead of a specific UV extension of the SM because we can thus investigate aspects common to a broad
class of models, which should make this study more relevant. It has previously been demonstrated~\cite{Gies:2013fua,Gies:2014xha,Eichhorn:2015kea},
using the functional renormalization group~(FRG) and various simplified versions of the SM, that the Higgs lower mass bound can indeed
be lowered when higher dimension operators are included.

Since the top Yukawa coupling is of order one, and because we want to study the EW finite temperature transition, it is desirable
to use a nonperturbative approach, i.e. a lattice regularization of the model; for a perturbative study, see~\cite{Grojean:2004xa}.
Unfortunately, it is not known how to regularize chiral fermion interactions on the lattice so we cannot consider the full SM.
There are two sectors of the SM which can be studied separately, the gauge-Higgs sector, consisting of the weak gauge bosons and
the Higgs field, and the Higgs-Yukawa sector, consisting of the Higgs field and the SM fermions. For a study of the first,
see~\cite{Kajantie:1996mn,Csikor:1998eu,Aoki:1999fi,Steinbauer:mt,Steinbauer:2015}. Here, we choose the latter because of the large
contribution to the SM Higgs sector from the Higgs-top interaction. The method we will employ to study this model is called
extended mean field theory (EMFT), which has proven to be a highly accurate approximation when applied to scalar field
theories~\cite{Akerlund:2013fsa,Akerlund:2014mea}. This allows us to obtain results with a computing effort orders of magnitude
smaller than with full Monte Carlo simulations of the same model~\cite{Bulava:2012rb,Hegde:2013mks,Chu:2015nha}, in addition to
other advantages explained in the main body of this paper. See also~\cite{Bulava:2013ep} for a study of the Higgs-Yukawa model
with an additional heavy fourth fermion family.

The rest of the paper is organized as follows. In section~\ref{sec:hym} we introduce the Higgs-Yukawa model and its lattice
discretization followed by details on the implementation of the chiral fermions in section~\ref{sec:diag_ov}. Section~\ref{sec:emft}
is devoted to the effective action in the EMFT approximation followed by our main results in section~\ref{sec:results}. We conclude in section~\ref{sec:conclusions}.

\section{Higgs-Yukawa model}\label{sec:hym}\noindent
The Higgs-Yukawa model is a simplified version of the SM Higgs sector where
the gauge degrees of freedom are neglected. The components of the model are the scalar complex doublet $\varphi$
and the fermion doublets $\Psi_f$. These couple to $\varphi$ via Yukawa couplings with coupling constants $y_f$,
which also determine the tree level fermions masses $m_f = y_fv$ after symmetry breaking, via the Higgs
field expectation value $\expv{\varphi}\equiv(0,v)^\intercal$. Since the top quark is orders of magnitude heavier than the other fermions,
it is common, and well justified, to restrict the fermion content to solely the top-bottom doublet
$\Psi= (t,b)^\intercal = (t_L,t_R,b_L,b_R)^\intercal$. In these fields the Euclidean continuum action is given by
\begin{equation}\label{eq:action_cont}
S^\text{cont}[\xbar{\Psi},\Psi,\varphi] = S_\text{H}[\varphi] + S_\text{F}[\xbar{\Psi},\Psi,\varphi],
\end{equation}
with
\begin{align}
  S_\text{H}[\varphi] &= \!\int\!\!\rd^4x\,\left\{\frac{1}{2}\abs{\partial_\mu\varphi}^2 + \frac{1}{2}m_0^2\abs{\varphi}^2
    +\hat\lambda\abs{\varphi}^4\right\},\\
  S_\text{F}[\xbar{\Psi},\Psi,\varphi] &= \!\int\!\!\rd^4x\,\left\{\xbar{\Psi}\slashed{\partial}\Psi + y_b\xbar{\Psi}_L\varphi b_R
    +y_t\xbar{\Psi}_L\widetilde{\varphi}t_R + \text{h.c.}\right\},
\end{align}
where $\widetilde{\varphi} = i\tau_2\varphi^\dagger$, $\tau_2$ is the second Pauli matrix and $\Psi_{L/R}=\left(t_{L/R},b_{L/R}\right)^\intercal$. The other fermions (quarks and leptons) can,
if desired, be added in a completely analogous way.

In this study we will mainly be interested in the symmetry broken phase and for notational convenience we will exploit the global
SU(2) symmetry to make the expectation value of $\varphi$ real and sit entirely in the lower component of $\varphi$,
i.e. we parametrize
\begin{equation}
  \label{eq:phi_param}
  \varphi(x) = \begin{pmatrix} g_2(x)+ig_1(x)\\v + h(x) -ig_3(x)\end{pmatrix},\; \expv{\varphi} = \begin{pmatrix} 0\\v\end{pmatrix},
\end{equation}
where $v+h(x)$ is the Higgs field and $g_i(x)$ are the three Nambu-Goldstone modes.

For the Higgs self-interaction we will consider higher dimension operators, in addition to the renormalizable $\phi^4$ interaction.
The simplest extension is to add a dimension six contact term with six Higgs fields, $(\varphi^\dagger \varphi)^3$, as studied by FRG
in~\cite{Eichhorn:2015kea}, but operators of any dimension could just as well be included. Let us for now simply group these higher
order terms in a ``new physics'' action~\footnote{We consider here only operators in the pure Higgs sector, for a complete list of
  dimension 6 operators see~\cite{Grzadkowski:2010es}}
\begin{equation}
  \label{eq:act_np}
  S_\text{NP}[\varphi] = \sum_{d=6}^{d_\text{max}}\sum_{i=1}^{n_d}C_{i,d}\frac{\mathcal{O}^i_d}{M_\text{BSM}^{d-4}},
\end{equation}
where $\mathcal{O}^i_d$ is an operator of mass dimension $d$, $n_d$ is the number of operators with dimension $d$, $C_{i,d}$ are the Wilson coefficients and
$M_\text{BSM}$ is the energy scale of the new physics, typically the mass of the lightest mediator particle.
Apart from $\varphi^6$ there is a second operator of dimension six,
$\mathcal{O}^2_6=\abs{\partial_\mu\varphi^\dagger\varphi}^2$, but it can be neglected if one assumes an approximate custodial
symmetry~\cite{Sikivie:1980hm}. Naturally (naively), at low energy $E$, the effects of the higher-dimension operators
are suppressed by factors $(E/M_\text{BSM})^{d-4}$.

\section{Diagonalizing the Overlap operator for arbitrary constant Higgs field}\label{sec:diag_ov}\noindent
To efficiently, albeit approximately, integrate out the fermions we consider the Higgs field to be very slowly varying
in space-time. Since we are primarily interested in the infrared properties of the model this assumption is reasonable.
We thus assume that the fermions see a constant Higgs field, therefore the fermionic interaction can be diagonalized by going
to Fourier space and the fermion determinant can be calculated without much effort.

Due to the global SU(2) invariance, the fermion determinant can only depend on the magnitude of the Higgs field,
$\abs{\varphi}^2= (v+h)^2+g_1^2+g_2^2+g_3^2$. Note that it depends on all the fields in
$\varphi$ and not only on the expectation value. To simplify the derivation we apply an SU(2) transformation such that
$\varphi = (0,\abs{\varphi})^\intercal$. Then, the different fermion flavors decouple and we have 
\begin{align}
S_\text{F} &= \sum_fN_{c,f}\int\rd^4x\, \bar{f}M_ff,\\
M_f &\equiv \slashed \partial + y_f\abs{\varphi}\bm{I}_4,\label{eq:fermop}
\end{align}
where $N_{c,f}$ is the number of colors for each fermion $f$, i.e. one for the leptons and three for the quarks. Unless otherwise
specified we will include all SM fermions except the neutrinos with their Yukawa couplings set via the tree level relation $y_f = m_f/v$.

For the model to be a realistic approximation to the SM Higgs sector it is important that the fermions
be chiral. This is ensured by implementing the Neuberger overlap operator~\cite{Neuberger:1997fp} when putting the fermions on the lattice.
The overlap operator satisfies an exact lattice chiral symmetry which approaches the usual chiral symmetry
in the continuum limit $a\to0$, $a$ being the lattice spacing.
Since we work with an effective model with a finite cutoff, this term will
never completely go away but as long as the cutoff is well above the top mass the effects should be small.
The overlap operator is given by
\begin{equation}
D^{(\text{ov})} = \frac{\rho}{a}\left(\bm{I}_4+\frac{A}{\sqrt{A^\dagger A}}\right),\;A=D^{(\text{W})}-\frac{\rho}{a},\;0<\rho\leq 2r,
\end{equation}
where $D^{(\text{W})}$ is the usual Wilson operator with negative bare mass $M_0$ and Wilson parameter $r$ and $\rho$ is a dimensionless parameter.
In our calculations we will adopt the common choices of $r=1/2,\,\rho=1$. The lattice action is constructed by the
following replacements:
\begin{align}
\slashed \partial &\to D^{(\text{ov})},\; \bar{f}_{L,R}f_{R,L} = \bar{f}P_{R,L}\hat{P}_{R,L}f,\\
\hat{P}_{R,L} &= \frac{\bm{I}_4\pm\gamma_5(\bm{I}_4-\rho^{-1}D^{(\text{ov})})}{2} = P_{R,L} \mp \frac{\gamma_5}{2\rho}D^{(\text{ov})},
\end{align}
after which the fermion operator, Eq.~\eqref{eq:fermop}, becomes
\begin{equation}
  \label{eq:fermop_ov}
  M_f^{(\text{ov})} = D^{(\text{ov})} + y_f\abs{\varphi}\left(\bm{I}_4 - \frac{1}{2\rho}D^{(\text{ov})}\right).
\end{equation}

In order to determine the fermion contribution to the action we will have to calculate the determinant, or equivalently the
trace log, of this operator. This is most convenient in Fourier space where $D^{(\text{ov})}$ is diagonal.
For a given $4$-momentum $p$, its four
eigenvalues come as complex conjugate pairs $\nu(p),\nu^\dagger(p)$, each with multiplicity two, where
\begin{align}
  \nu(p) &= \rho\left(1+\frac{i\sqrt{\tilde{p}^2}+\frac{r}{2}\hat{p}^2-\rho}{\sqrt{\tilde{p}^2+\left(\frac{r}{2}\hat{p}^2-\rho\right)^2}}\right),\\
\tilde{p}^2 &= \sum_\mu\sin^2(p_\mu),\; \hat{p}^2 = 4\sum_\mu\sin^2\left(\frac{p_\mu}{2}\right).
\end{align}
Since the ``mass term'' $y_f\abs{\varphi}$ is real, the determinant of $M_f^{(\text{ov})}$ is real as well and the trace log takes
the form of a real integral
\begin{align}
  \text{TrLog}&\left(M_f^{(\text{ov})}\right) \\
  &= 2\!\!\int\!\!\frac{\rd^4p}{(2\pi)^4}\log\abs{\nu(p) + y_f\abs{\varphi}\left(1 - \frac{\nu(p)}{2\rho}\right)}^2,\nonumber
\end{align}
which can be calculated quite efficiently. Actually, since it only depends on one variable, $y_f\abs{\varphi}$,
and will have to be evaluated very often, it will prove advantageous to precalculate it on a discrete set of values and interpolate to
intermediate points. In summary, to a first approximation the effect of the fermions is the addition of an SU(2)
symmetric contact term to the Higgs potential.

It should be noted that the determinant is only positive for generic Yukawa couplings in the approximation that the
Higgs field is constant. Otherwise the Higgs field will fluctuate in the complex plane and introduce a sign problem,
unless the fermions in each doublet have degenerate Yukawa couplings.

\section{The effective action and EMFT solution}\label{sec:emft}\noindent
For definiteness we will consider only the $\abs{\phi}^6$ term in $S_\text{NP}$, Eq.~\eqref{eq:act_np}. Since we have no handle on the Wilson coefficient of this
term we will set it to $1$ and introduce $\lambda_6 \equiv (aM_\text{BSM})^{-2}$ where $a$ is the lattice spacing.
With the approximate treatment of the fermions above, we end up with the lattice action
\begin{widetext}
\begin{equation}
  S[\varphi] = \sum_x\Bigg\{-\kappa\sum_\mu\varphi^\dagger_x\varphi_{x+\hat\mu}+\text{h.c.} + \abs{\varphi_x}^2 +
  \hat{\lambda}\left(\abs{\varphi_x}^2-1\right)^2 + \sum_fN_{c,f}\text{TrLog}\left(M_f(y_f\sqrt{2\kappa}\abs{\varphi_0})\right) + \hat{\lambda}_6\abs{\phi}^6\Bigg\},  \label{eq:action_trlog}
\end{equation}
\end{widetext}
in terms of the conventional $\varphi^4$ parameters
\begin{align}
  a\varphi(x) &= \sqrt{2\kappa}\varphi_x,\;(am_0)^2=\frac{1-2\hat{\lambda}}{\kappa}-8\\
  \hat\lambda &= 4\kappa^2\lambda,\;\hat{\lambda}_6 = 8\kappa^3\lambda_6.\nonumber
\end{align}
This action is quite similar to the complex
$\varphi^4$ model studied by EMFT in \cite{Akerlund:2014mea} and we can adopt the vector notation used there,
\begin{equation}
  \bphi_x^\intercal = (h_x,g_{1,x},g_{2,x},g_{3,x})^\intercal+(\hat{v},0,0,0)^\intercal\equiv \bdphi_x^\intercal+\expv{\bphi}^\intercal,
\end{equation}
with $\hat{v} = av/\sqrt{2\kappa}$, to derive the EMFT equations in an analogous fashion. Concentrating on the field at
the origin, $\bphi_0$, the hopping part of the action can be expressed as
\begin{equation}
  \label{eq:hopping}
  \Delta S = -2\kappa\sum_{\pm\mu}\bdphi_0^\intercal\bdphi_{\hat{\mu}} -4d\kappa \hat{v} h_0.
\end{equation}
The lattice without the origin is considered an external bath and will be self-consistently integrated out.
This is equivalent to replacing the nearest-neighbor interaction term in Eq.~\eqref{eq:hopping} by its cumulant
expansion with respect to the external bath. Truncating the expansion at second order we obtain the following effective
action:
\begin{align}
  S_\text{EMFT} &= \bphi^\intercal(\bm{I}-\bm{\Delta})\bphi + \hat\lambda\left(\norm{\bphi}^2-1\right)^2 \nonumber\\
  &+\sum_fN_{c,f}\text{TrLog}\left(M_f^{(\text{ov})}\left(y_f\sqrt{2\kappa}\norm{\bphi}\right)\right)\label{eq:emft_action}\\
  &- 2\hat{v}(\hat{v}+h)(2d\kappa-\Delta_1)+ \hat{\lambda}_6\norm{\bphi}^3,  \nonumber
\end{align}
where $\bm{\Delta}$ emulates propagation in the external bath. A closer inspection of the cumulant expansion reveals
\begin{align}
  \bm{\Delta} &= 2\kappa^2\sum_{\pm\mu,\rho}\expv{\bdphi_\mu\bdphi_\rho^\intercal}_\text{ext}\nonumber\\
  &=2\kappa^2\sum_{\pm\mu,\rho}\text{diag}\expv{\left(h_\mu h_\rho,g_{1,\mu} g_{1,\rho},g_{2,\mu} g_{2,\rho},g_{3,\mu} g_{3,\rho}\right)}_\text{ext}\nonumber\\
  &\equiv\text{diag}(\Delta_1,\Delta_2,\Delta_2,\Delta_2),\label{eq:delta}
\end{align}
where the diagonal form follows from the symmetries of the action. Since the action is still symmetric with respect to
O(3) rotations of $g_1,g_2,g_3$ it is practical to rewrite the action in terms of two variables
\begin{align}
  (\hat{v}+h_0) &= \phi_h, \\
  \sqrt{g_{1,0}^2+g_{2,0}^2+g_{3,0}^2} &= \phi_g,
\end{align}
in which $\Delta = \text{diag}(\Delta_1,\Delta_2)$ and $\expv{g_{i,0}^2} = \expv{\phi_g^2}/3$. The EMFT action is then
given by
\begin{widetext}
\begin{align}
  S_\text{EMFT} &= (1-\Delta_1)\phi_h^2 + (1-\Delta_2)\phi_g^2 + \hat{\lambda}\left(\phi_h^2+\phi_g^2-1\right)^2
  +\sum_fN_{c,f}\text{TrLog}\left(M_f^{(\text{ov})}\left(y_f\sqrt{2\kappa}\sqrt{\phi_h^2+\phi_g^2}\right)\right)\label{eq:emft_action_2}\\
  &\phantom{=}- 2\hat{v}\phi_h(2d\kappa-\Delta_1)+ \hat{\lambda_6}\left(\phi_h^2+\phi_g^2\right)^3,\nonumber
\end{align}
\end{widetext}
and the partition function becomes
\begin{equation}
  \label{eq:z_emft}
  Z_\text{EMFT} = \mathcal{N}\int\!\!\rd\phi_h\rd\phi_g\,\phi_g^2\exp\left(-S_\text{EMFT}\right),
\end{equation}
where $\mathcal{N}$ is an irrelevant normalization constant. The unknown parameters $v$ and $\bm \Delta$ can then
self-consistently be determined via the three self-consistency equations
\begin{align}
  \expv{\phi_h} &= \hat{v},\\
  2\expv{\phi_h^2}_c &= \!\!\int\!\!\!\frac{\rd^4p}{(2\pi)^4}\frac{1}{\frac{1}{2\expv{\phi_h^2}_c}+\Delta_1 - 2\kappa Z_h\sum_\mu\cos(p_\mu)},\\
  \frac{2\expv{\phi_g^2}}{3} &= \!\!\int\!\!\!\frac{\rd^4p}{(2\pi)^4}\frac{1}{\frac{3}{2\expv{\phi_g^2}}+\Delta_2 - 2\kappa Z_h\sum_\mu\cos(p_\mu)},
\end{align}
where $\expv{\phi_h^2}_c = \expv{\phi_h^2}-\expv{\phi_h}^2$ and the wave function renormalization $Z_h$ is chosen such that the Nambu-Goldstone
bosons are exactly massless~\cite{Akerlund:2014mea}.
The last two equations enforce that the connected $2$-point function, from the origin to the origin, is equal to its momentum-space expression.
The four- (or $d$-) dimensional integrals can be transformed into one-dimensional integrals by using the identity
\begin{equation}
  \int\!\!\frac{\rd^dp}{(2\pi)^d}\frac{1}{a-\sum_\mu\cos(p_\mu)} = \int\limits_0^\infty\!\!\rd\tau e^{-a\tau}\left(I_0(\tau)\right)^d,
\end{equation}
where $I_0(x)$ is a modified Bessel function of the first kind. See \cite{Akerlund:2014mea} for more details.

\subsection{Scale setting and observables}
In order to examine the effect of different cutoffs and possible higher dimension operators on the Higgs boson mass we
need to set the scale of the lattice calculations. This is most naturally done by fixing the Higgs field expectation value $v$
to its phenomenological value. In terms of lattice variables we have
\begin{equation}
  \label{eq:scale}
  v = \frac{\sqrt{2\kappa}\hat{v}}{a} = 246 \,\text{GeV},
\end{equation}
which, given $\hat{v}$, determines the value of the cutoff $\Lambda=1/a$. Furthermore, we want to use physical quark masses, so we fix the
Yukawa couplings using the tree level relation $m_f = y_fv$. We can now determine the Higgs boson mass $M_h$ in GeV as
a function of the parameters of the Higgs potential by evaluating the inverse propagator at zero momentum:
\begin{align}
  G_h^{-1}(p) &= \frac{1}{2\expv{\phi_h^2}_c}+\Delta_1-2\kappa Z_h\sum_\mu\cos(p_\mu)\nonumber\\
  &\underset{p\to 0}{\to} \kappa Z_h((aM_h)^2 + (ap)^2), \label{eq:higgs_mass}\\
  \Rightarrow M_h^2 &= \left(\frac{1}{2\expv{\phi_h^2}_c}+\Delta_1-8\kappa Z_h\right)\frac{\Lambda^2}{\kappa Z_h}. \nonumber
\end{align}

\section{Results}\label{sec:results}
\begin{figure}[htp]
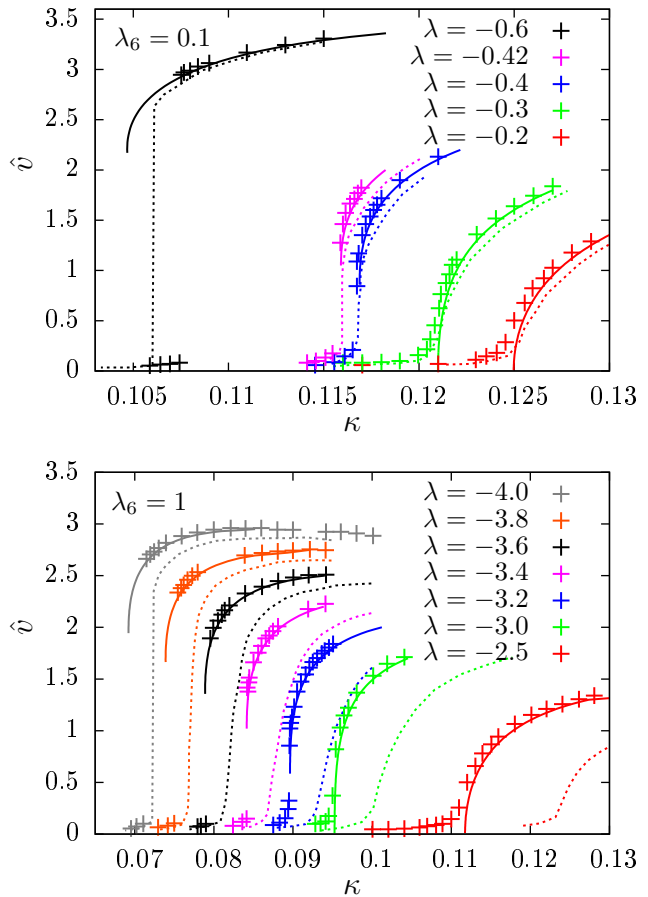

\centering
\includegraphics[width=1\linewidth]{{{figures/compare_l6_01}}}\\
\includegraphics[width=1\linewidth]{{{figures/compare_l6_1}}}
\caption{The Higgs vacuum expectation value in lattice units as a function of the coupling $\kappa$ at various $\lambda$, for $\lambda_6=0.1$
  (\emph{upper panel}) and $\lambda_6 = 1$ (\emph{lower panel}). The solid lines are EMFT calculations from this work and
  the pluses and dashed lines are full Monte Carlo simulations and CEP calculations respectively, both taken from~\cite{Hegde:2013mks}.
  The EMFT results follow the Monte Carlo data closely for both values of $\lambda_6$, whereas the CEP calculation gives reasonably accurate
  results for the upper, perturbative value only.}
  \label{fig:comparison}
\end{figure}

\noindent
In order to assess how well our EMFT method works in the presence of fermions, we compare it to already existing full Monte Carlo
results and results obtained using an analytic, approximate method, the constraint effective potential (CEP)~\cite{Chu:2015nha}.
In Fig.~\ref{fig:comparison} we show the Higgs expectation value in lattice units $\hat v$ as
a function of the hopping parameter $\kappa$ at two different values of the new coupling,
$\lambda_6=0.1$ in the upper panel and $\lambda_6=1.0$ in the lower one. Each color represents a different value of
the quartic coupling $\lambda$. In this comparison only the top and bottom quarks are included with degenerate
Yukawa couplings $y_b=y_t=175/246$ and color factors $N_{c,b}=N_{c,t}=1$. It is clear that EMFT (solid lines) is a very good approximation and gives results close to
the Monte Carlo results (symbols) in all cases, in contrast to CEP (dotted lines) which works acceptably well for small values
of $\lambda_6$ only. This is not surprising since the CEP calculations in~\cite{Hegde:2013mks} are perturbative, whereas EMFT is fully nonperturbative.

While we have no explanation for the remarkable accuracy of our EMFT approximation, we point out that a similar accuracy has been observed in the $\varphi^4$
lattice model near criticality~\cite{Akerlund:2013fsa}: EMFT appears to perform well near a Gaussian critical point.

For our actual results we will adopt a slightly different point of view on the model than the authors of~\cite{Chu:2015nha}.
Indeed, consider the origin of the higher dimension operators. Generically they stem from UV completion of the SM and
are thus associated with an energy scale which is typically the mass of the lightest of the ``new'' particles which couple
to the Higgs field. Let us call this energy scale $M_{\rm{BSM}}$, where \emph{BSM} stands for ``beyond the standard model.''
It is natural to assume that the coupling of this particle to the Higgs field
is of order one, such that the coefficient in front of the $\abs{\phi}^6$ operator will be $M_{\rm{BSM}}^{-2}$ with the corresponding
dimensionless coupling $\lambda_6 = (aM_\text{BSM})^{-2}$. Notice, moreover,
that $\Lambda=a^{-1}$, not $M_\text{BSM}$, is the cutoff of the effective model, since it is directly related to the maximum energy
scale probed by the lattice action. In order to justify the effective treatment of particles heavier than $M_\text{BSM}$ the cutoff $\Lambda$
has to be sufficiently small. This leads to a hierarchy of scales condition, $aM_h\ll 1 \lesssim aM_{\rm{BSM}}$, which in turn means $\lambda_6\lesssim 1$
in the broken symmetry phase. As $(aM_\text{BSM})$ is decreased toward $1$, more and more terms in the effective action would have to be taken into
account in order to maintain a good approximation of the underlying model.

\begin{figure}[htp]
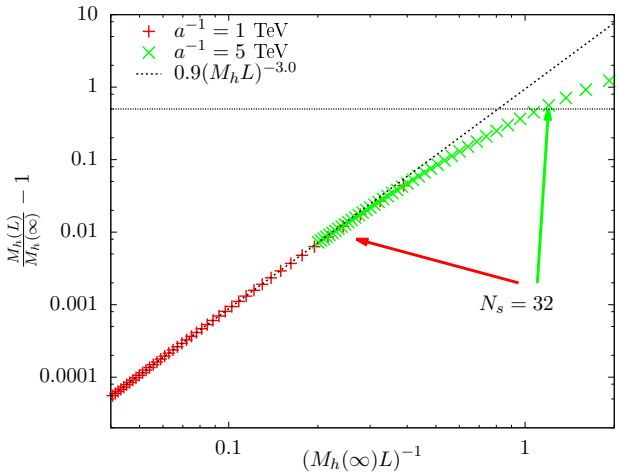

\centering
\includegraphics[width=1\linewidth]{{{figures/fv_l6_01}}}
\caption{Finite volume relative correction to the Higgs mass for two different lattice spacings calculated with EMFT with
  $\lambda_6 = 1/10$ and $M_h=\xi^{-1}=125$ GeV. The horizontal line corresponds to a $50\%$ correction. In order to see the
  asymptotic $\left(\xi/L\right)^3$ corrections due to the massless Nambu-Goldstone modes, rather large lattices are needed,
  which poses a challenge to full Monte Carlo simulations. In EMFT one can avoid the problem of thermodynamic extrapolation
  entirely by working directly in the thermodynamic limit.}
\label{fig:fin_vol}
\end{figure}

It is quite challenging to preserve a good separation of scales while at the same time keeping the physical volume large, and hence any lattice
simulation is susceptible to large finite size effects. This is particularly true in a theory with massless modes, like the one we study here.
In EMFT one generally works directly in the thermodynamic limit but in order to demonstrate the power law corrections coming from
the Nambu-Goldstone modes, we have solved the self-consistency equations in a finite volume. In Fig.~\ref{fig:fin_vol} we show the
relative error on the Higgs mass as a function of the box size. The scale separation factor is
$aM_{\rm{BSM}}=\sqrt{10}$ ($\lambda_6=0.1$) for all values of $a$ and we have marked both where the size of the correction is $50\%$ of the mass itself and
where an $N_s=32$ lattice would be for two different lattice spacings. This demonstrates the need for very big lattices before one can
even see the asymptotic power law scaling.

\subsection{The zero-temperature phase diagram}
\begin{figure}[htp]
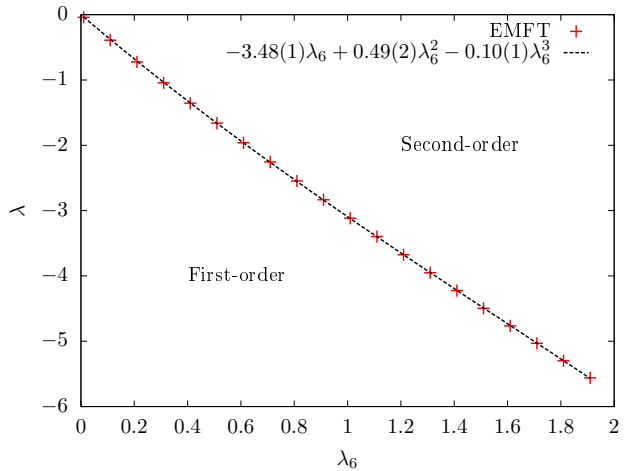

\centering
\includegraphics[width=1\linewidth]{{{figures/first_order_line}}}
\caption{The tricritical point at zero temperature. For $\lambda$ below the line the transition is first order and above
  it is second order.}
  \label{fig:first_order_line}
\end{figure}

\begin{figure}[htp]
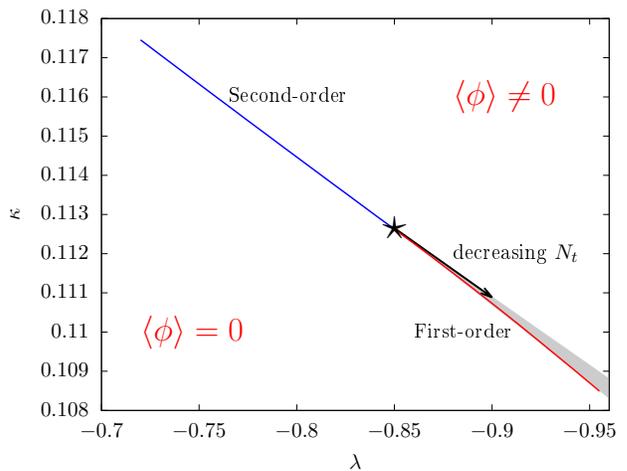

\centering
\includegraphics[width=1\linewidth]{{{figures/pb_l6_4}}}
\caption{Zero- and finite-temperature transition in the $(\lambda,\kappa)$-plane at fixed $\lambda_6=1/4$. The transition turns from
  second to first order at the first-order endpoint marked by the star. The color of the line denotes the order of the transition, blue for second order and red for first order.
  As the lattice size is reduced in the temporal direction the endpoint moves along the arrow, and thus the gray shaded area marks the region in the plane
  where the finite temperature transition is first order.}
  \label{fig:pb_l6_4}
\end{figure}

\noindent
It is most convenient to present the phase diagram in the (unphysical) bare parameters $\kappa,\lambda$ and $\lambda_6$. One can
then pass to physical units via the Higgs expectation value in the broken phase. In Fig.~\ref{fig:comparison} one can see that the transition
turns from second- to first-order as $\lambda$ is made more negative and in Fig.~\ref{fig:first_order_line} we show how the tricritical point depends on $\lambda_6$. For $\lambda$ below the line the $\kappa$-driven transition is first order and above it is second order~\footnote{In the absence of gauge fields there is always a symmetric phase and a broken phase separated by a phase transition, never a crossover}.
Next, we fix $\lambda_6$ and look at the transition in the $(\lambda,\kappa)$-plane. An example, where $\lambda_6=1/4$, can be seen in
Fig.~\ref{fig:pb_l6_4}. The color of the line denotes the order of the transition, blue for second order and red for first order.
The star marks the tricritical point and the arrow denotes how it moves as the number of lattice sites
in the temporal direction $N_t$ is decreased (see below for more details). The location of the tricritical point is obtained by fitting
the critical vev on the first-order side with a power law $\expv{\phi}_c(\lambda) = c \left(\lambda_c-\lambda\right)^b$.
There is a region close to the second-order line where one can perform calculations at a small
lattice spacing $a$ and since it is not possible to take the continuum limit of this effective theory anyway,
it may also be viable to stay close to the transition on the first-order side. In fact, it turns out that a first order finite
temperature transition will be found only there, denoted by the gray shaded area in the figure.

\subsection{Higgs mass lower bound}\label{sec:mass_bound}\noindent
Given a specific form of the Higgs potential, the lower bound on the Higgs mass is simply given by the minimal mass obtainable
in the given parameter space of the potential~\footnote{Note that the effective potential can feature local minima,
which may have a smaller curvature than the global one. We are not interested in those.}. For a pure $\varphi^4$ potential it is found that the Higgs mass decreases
when $\lambda$ decreases and the lower bound is thus obtained at vanishing quartic coupling. In the Higgs-Yukawa model, this lower bound
turns out to be just above $40$ GeV and a positive $\lambda$ is needed to bring the Higgs mass up to $125$ GeV. A negative coupling is
obviously prohibited by the requirement of a bounded action. By introducing higher dimension operators, we can have a
stable vacuum even at negative quartic coupling, and it is plausible that this could lead to an even lower Higgs mass. This was
first demonstrated in~\cite{Gies:2013fua}, using the FRG on a chiral $\Z_2$ Higgs-Yukawa model. Analogous results were also obtained using
the Higgs-Yukawa model described above at physical values of the top and bottom masses in~\cite{Gies:2014xha} and using the chiral $\Z_2$
Higgs-Yukawa model plus an $SU(3)$ gauge sector in~\cite{Eichhorn:2015kea}. Later, the authors of~\cite{Chu:2015nha}
came to similar conclusions using nonperturbative Monte Carlo simulations and perturbative CEP calculations of the above described Higgs-Yukawa model
with mass-degenerate top and bottom masses. Common to all these studies is that they add a $\phi^6$ operator to the Higgs potential and when
its coupling constant $\lambda_6$ is positive, the Higgs mass can be further reduced by making the quartic coupling $\lambda$
more and more negative. At some point, in the lattice regularized models, the phase transition between the symmetric and broken phases turns first
order (see Fig.~\ref{fig:first_order_line}) and there is a hard lower bound on the lattice spacing $a$. Since the model is only effective
this is in itself not a problem, but, since one wants $aM_h\ll1$, it bounds the region in parameter space where simulations are useful.

\begin{figure}[htp]
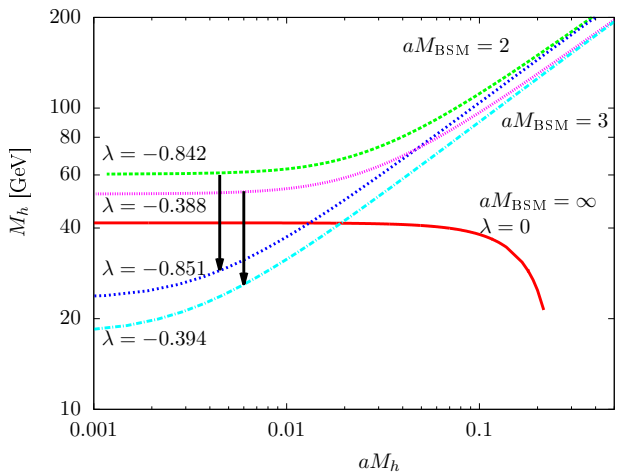

\centering
\includegraphics[width=1\linewidth]{{{figures/lower_mass}}}
\caption{The Higgs mass $M_h$ as a function of $aM_h$ for various combinations of $aM_\text{BSM}$ and $\lambda$, together with the SM lower bound
  on the Higgs mass obtained at $M_\text{BSM}=\infty$ and $\lambda=0$. In the presence of a $\phi^6$ operator a negative value of $\lambda$ is
  allowed and the Higgs mass can be lowered well below the SM bound, as indicated by the arrows.}
  \label{fig:lower_mass}
\end{figure}

Typically one finds that the Higgs mass goes to zero as one approaches the tricritical point from the second-order side although
before zero is obtained one runs into subtle issues regarding new local minima of the effective action~\footnote{This scenario is however
  quite tricky to study in an effective model setup since the higher field value at the new minimum reduces the separation of
  the dynamical and the cutoff scales which might make the effective model less precise}. This means that the Higgs mass can in
general be lowered by a large if not arbitrary amount, from its lower bound in the $\lambda_6=0$ case. This is demonstrated in
Fig.~\ref{fig:lower_mass}, where the Higgs mass $M_h$ is plotted as a function of $aM_h$ for a few different values of $aM_\text{BSM}$
and $\lambda$, together with the SM lower bound obtained at $\lambda=0$. As $\lambda$ is made more negative one approaches the regime of
first-order transition and the Higgs mass decreases and can take values well below the SM lower bound.

All in all it is clear that in the presence of higher dimension operators
the lower bound on the Higgs mass loses its meaning. In fact, one may argue that this is a null statement since the lower bound was
calculated under the assumption that there is only the SM, and clearly new operators will change the running of the couplings.

\subsection{Finite temperature}\noindent
Let us now turn to the $M_\text{BSM}$ dependence of the finite temperature transition. In gauge-Higgs systems it has been demonstrated
that introducing a $\varphi^6$ operator makes the phase transition stronger~\cite{Grojean:2004xa,Steinbauer:mt,Steinbauer:2015}, which in turn means that
the critical mass, up to which the transition is first order, increases. If it would increase past the observed Higgs boson mass, electroweak baryogenesis might become possible again. Here, we present our findings using EMFT. Still in the infinite volume limit, we can introduce
a nonzero temperature $T=1/(aN_t)$ by using a finite number $N_t$ of lattice points in the temporal direction. This gives us control over the temperature
in discrete steps (for a fixed lattice spacing), so in order to get a good resolution one would need to work with rather fine lattices.
However, this limits the range of available $M_\text{BSM}$ because of the condition of scale separation $aM_\text{BSM}\gtrsim 1$. To overcome this problem we will
determine the linear response of the system and then extrapolate to the desired temperature. Alternatively, to continuously vary the
temperature, one could use a lattice action with anisotropic couplings.

\begin{figure}[htp]
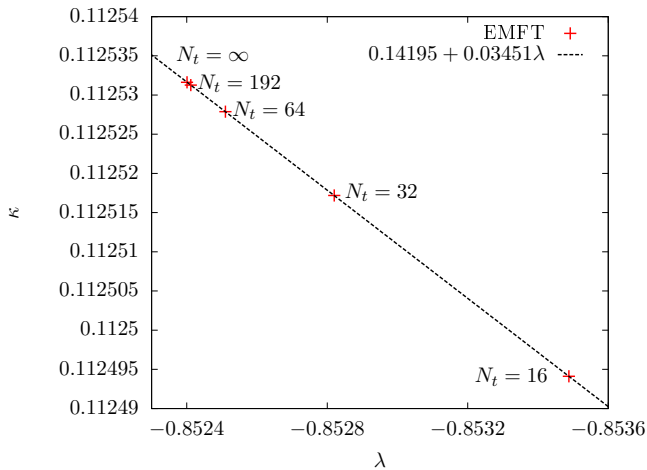

\centering
\includegraphics[width=1\linewidth]{{{figures/endpoints_l6_4}}}
\caption{The $N_t$ dependence of the tricritical point for $\lambda_6=1/4$. The trajectory is very well described by a straight line.}
  \label{fig:endpoints_l6_4}
\end{figure}

\begin{figure}[htp]
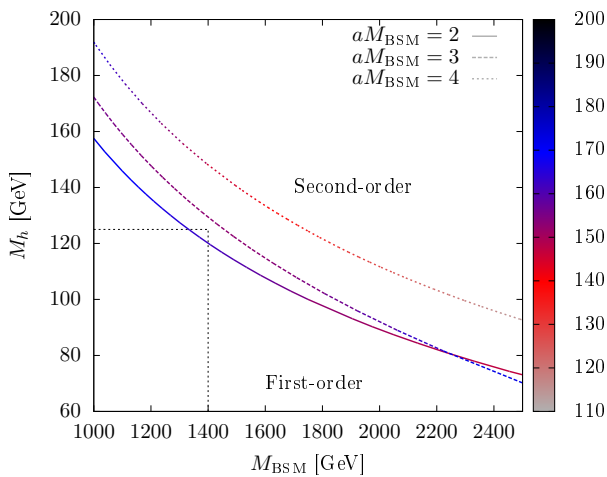

\centering
\includegraphics[width=1\linewidth]{{{figures/endpoint_ferm}}}
\caption{The critical Higgs mass below which the finite temperature transition is first order for various values of $aM_\text{BSM}$.
  The color coding gives the transition temperature in GeV. The 3 curves give  a measure of the sensitivity of our effective theory to the cutoff.
  A higher cutoff makes the transition weaker.}
  \label{fig:crit_mass}
\end{figure}

\begin{figure}[htp]
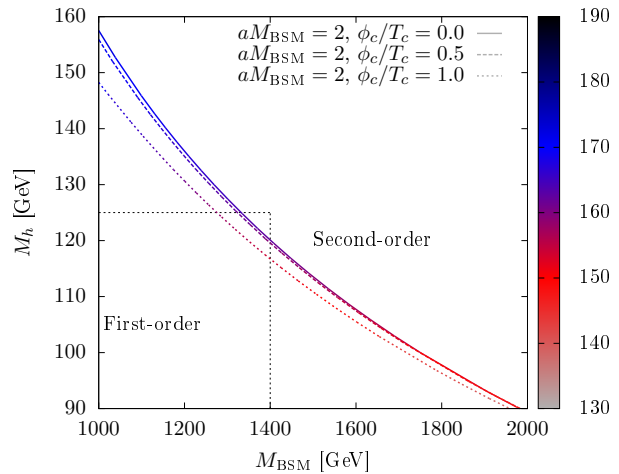

\centering
\includegraphics[width=1\linewidth]{{{figures/endpoint_str_lam6_025}}}
\caption{The critical Higgs mass below which the finite temperature transition is first order for $aM_\text{BSM}=2$ and different strengths of
  the transition measured in terms of $\phi_c/T_c$ where $\phi_c$ is the critical Higgs vev at the transition.
  The color coding gives the transition temperature in GeV.}
  \label{fig:crit_mass_str}
\end{figure}

The observables of main interest are the critical Higgs mass for which the transition turns first order and the critical temperature.
Another interesting observable is the Higgs mass for which the vev at the transition is of the same order as the critical temperature,
which is the actual condition for a viable EW baryogenesis.
All observables depend on both $M_\text{BSM}$ and the lattice spacing $a$ so we need to calculate them in a two dimensional parameter
space. The starting point for determining this dependence is to obtain the phase diagram as in Fig.~\ref{fig:pb_l6_4} for various
$aM_\text{BSM}$ and $N_t$ values, and then to determine how the first-order endpoint moves as a function of $N_t$. This is illustrated by the star
and the arrow in Fig.~\ref{fig:pb_l6_4}. The finite temperature transition will be first order in the region between the $T=0$ first-order
line and the trajectory of the first-order endpoint, denoted by the shaded region in Fig.~\ref{fig:pb_l6_4}. We find that the endpoint
moves on a straight line in the $(\kappa,\lambda)$-plane, as can be seen in Fig.~\ref{fig:endpoints_l6_4} for $\lambda_6=1/4$.
By determining the lattice spacing $a$ and the Higgs mass $M_h$ at zero temperature along this line we can obtain the critical Higgs mass
as a function of the BSM scale, shown in Fig.~\ref{fig:crit_mass}.
In Fig.~\ref{fig:crit_mass_str} we show how the critical Higgs mass for $\lambda_6=1/4$ changes for different strengths of the first
order transition, measured in terms of $\phi_c/T_c$ where $\phi_c$ is the critical Higgs expectation value
at the phase transition. In both of these figures the color of the line gives the critical temperature $T_c$ in GeV. 

\FloatBarrier

Since our model is an effective one, we do not expect that the resulting curve is independent of the lattice spacing.
Indeed, in the range of $2\leq aM_\text{BSM}\leq 4$ we find that the critical Higgs mass varies by about ten percent. This is in itself not a direct measure
of the systematic uncertainties of the study since the choice of only including the $\varphi^6$ operator introduces uncertainties which are hard to quantify,
especially at the lower end of the interval. Moreover, different implementations of a UV cutoff will give somewhat different results. With these caveats,
the variation of the critical line in the $(M_h,M_\text{BSM})$-plane gives a measure of the uncertainties
within the model itself. The final result is that for a Higgs mass of $125$~GeV a BSM scale of around $1.5$~TeV is needed to make the EW finite temperature
phase transition first order, and this result changes only slightly even if we demand that the transition should be strongly first order with
$\phi_c/T_c\gtrsim 1$. This result is interesting since in many supersymmetric extensions of the SM one expects to find the lightest superpartners around
this mass scale. The final scale separations $(aM_h\ll1\ll{}aM_\text{BSM})$ at the tricritical point for the different values of $aM_\text{BSM}$ are listed in Table~\ref{tab:scales}
and it is evident that the validity of the effective model is somewhat strained, but not completely spoiled, due to the rather small separations.

\begin{table}[htp]
  \centering
  \caption{Separations of three scales in units of the inverse lattice spacing at the tricritical point for different values of $aM_\text{BSM}$.}
  \begin{tabular}{c c c}
    $M_\text{BSM}$ & $\quad{}a^{-1}\quad$ & $M_h$\\\hline
    $2$ & 1 & $0.19$\\
    $3$ & 1 & $0.26$\\
    $4$ & 1 & $0.29$
  \end{tabular}
  \label{tab:scales}
\end{table}

\begin{figure}[htp]
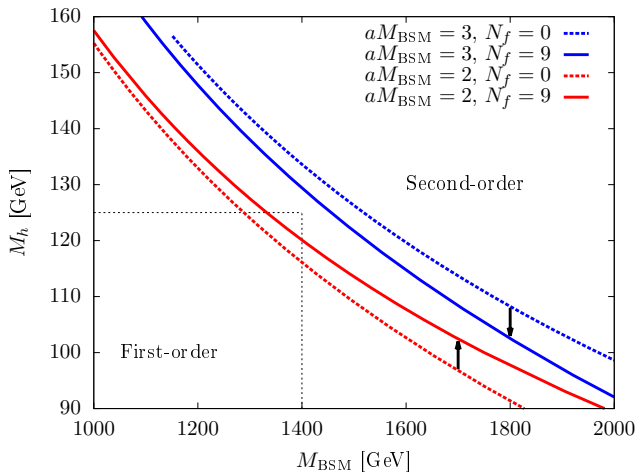

\centering
\includegraphics[width=1\linewidth]{{{figures/endpoint_noferm}}}
\caption{The critical Higgs mass below which the finite temperature transition is first order for various values of $aM_\text{BSM}$ with and
  without the SM fermions. The correction due to the fermions (shown by the arrows) is small and of indefinite sign.}
  \label{fig:crit_mass_ferm}
\end{figure}

Finally, to quantify the influence of our nonperturbative treatment of the fermions we repeated the calculations considering just the Higgs sector
and found that, after all, the fermions contribute only percent-level corrections to the purely bosonic case, see Fig.~\ref{fig:crit_mass_ferm}.
Remarkably, the sign of the correction depends on the value of $aM_\text{BSM}$. We also note that the BSM scale of $\approx 1.5$~TeV needed for
a first-order finite-temperature transition is in good agreement with what one obtains in the gauge-Higgs model with a
$\phi^6$-term~\cite{Steinbauer:mt,Steinbauer:2015}. This could be used as an argument for leaving out
the SM fermions and gauge fields from the simulations while studying higher dimension operators in the context of EW baryogenesis, and for
including them perturbatively only.

\section{Conclusions}\label{sec:conclusions}\noindent
We have demonstrated that EMFT agrees very well with full Monte Carlo simulations of the Higgs-Yukawa model which contain both
scalar fields and chiral fermions. We have then gone beyond what is possible for Monte Carlo simulations by lifting the mass degeneracy
of the top and bottom quarks as well as including all other SM fermions. We have furthermore studied the EW finite temperature
transition in the presence of a $\varphi^6$ term in the Higgs potential and thus obtained the critical Higgs mass for which this
transition turns first order, something which has not been done nonperturbatively before. We find that with a BSM scale of about $1.5$~TeV
the transition turns first order for a Higgs mass of $M_h=125$ GeV. At this point the effective model shows a separation of roughly
a factor $3$ between the relevant scales $M_h$, $a^{-1}$ and $M_\text{BSM}$. The value of $M_\text{BSM}$ decreases only slightly if we demand a
strong first order transition with $\phi_c/T_c\gtrsim1$. This scale is consistent with what is found in the gauge-Higgs
model~\cite{Steinbauer:mt,Steinbauer:2015}, where the EW gauge fields are taken into account but the fermions are neglected,
as well as in the perturbative study~\cite{Grojean:2004xa}. The scale is also only mildly dependent on the exact value of the lattice
cutoff $1/a$ within the window between $M_h$ and $M_\text{BSM}$. It is however difficult to assess the effect of neglecting other higher order
operators. We have further shown that removing the fermions altogether shifts the critical Higgs mass only by a few percent, establishing that
the Higgs sector itself is the dominating driving factor of the transition. To confirm that the gauge and fermion sectors always yield small contributions to
the critical mass it would be interesting to study a model with different higher dimension operators, in particular one including both the Higgs field and the
gauge fields.

\section*{Acknowledgments}
We thank KITP at UC Santa Barbara for its hospitality. This research was supported in part by the National Science Foundation 
under Grant No. NSF PHY11-25915.

\bibliography{hy_submit.bib}

\end{document}